\documentstyle[epsfig,float,twocolumn,prl,aps]{revtex}

\begin{document}
\draft

\twocolumn[\hsize\textwidth\columnwidth\hsize\csname @twocolumnfalse\endcsname

\title{Magnetic field effects on two-leg Heisenberg antiferromagnetic ladders:\\
Thermodynamic properties}
\author{ Xiaoqun Wang$^{1}$ and Lu Yu$^{2,3}$ }

\address{
$^1$Institut Romand de Recherche Numerique en Physique Des
Materiaux, PPH-333, EPFL, CH-1015 Lausanne, Switzerland
 }
\address{
$^2$International Center for Theoretical Physics, P. O. Box 586,
34100 Trieste, Italy
}
\address{
$^3$Institute of Theoretical Physics, P. O. Box 2735, Beijing
100080, P.R. China
}
\maketitle
\begin{abstract}
{Using the recently developed transfer-matrix renormalization group method, 
we  have studied the thermodynamic properties of two-leg 
antiferromagnetic ladders in the magnetic field. Based on  different  behavior of magnetization,
we found disordered spin liquid, Luttinger liquid, spin-polarized phases and a classical regime depending on
magnetic field and temperature. Our calculations  in Luttinger liquid regime suggest that
both  the divergence of the NMR relaxation rate and the anomalous specific heat behavior
observed on Cu$_2$(C$_5$H$_{12}$N$_2$)$_2$Cl$_4$} 
are due to quasi-one-dimensional effect rather than three-dimensional ordering.
\end{abstract}
]

Recently spin ladders have been at the  focus of intensive research 
towards understanding the spin-1/2 Heisenberg antiferromagnets 
in one and two dimensions\cite{Rice,dago,zhang,barnes,frust}.
Experimentally, several classes of  materials like SrCu$_2$O$_3$,
La$_6$Ca$_8$Cu$_{24}$O$_{41}$ and Cu$_2$(C$_5$H$_{12}$N$_2$)$_2$Cl$_4$ 
(CuHpCl)  have been found whose properties can be well described 
by the two-leg Heisenberg antiferromagnetic 
ladder (THAFL) model\cite{Az,imai,Chaboussant97a}. For inorganic oxides,  
the spin gap $\Delta$ was found around $500K$\cite{Johnston96}. So
only  the low-energy part of spectrum can be  explored by 
measuring   spin susceptibility,
 NMR relaxation and   neutron scattering\cite{Az,imai,Ecc}.
On the other hand, the organo-metallic compound CuHpCl exhibits a very small 
spin gap $\Delta$$\approx$$11K$\cite{Chaboussant98} which
 allows a full investigation of the spectrum
by applying a  magnetic field  (MF).  
Chaboussant  {\it et al.} have shown 
that the NMR relaxation rate exhibits substantially different behavior for 
different ranges of MF in the low temperature limit.  On this basis
these authors proposed a magnetic phase diagram\cite{Chaboussant98}.
Moreover, the specific heat measurements show anomalous behavior when
the spin gap is suppressed by the MF\cite{Hammar98,Cal}.
Theoretically, some of the MF effects on THAFL were discussed
by using  exact diagonalization, bosonization, conformal field theory and 
non-linear $\sigma$-model approaches
\cite{Cal,Hawyard96,chitra,usami,sakai,norman,mt}.
In this paper, we perform the first calculation of the phase diagram
(more precisely crossover lines between different regimes) using the newly developed
transfer-matrix renormalization group (TMRG) technique\cite{wang}.
Our findings suggest  the observed divergence 
of the NMR rate\cite{Chaboussant98} and the anomalous specific heat
behavior\cite{Hammar98,Cal} are  due to the MF effects on THAFL
{\it i.e.}, quasi-1D effects rather than  3D field-induced ordering when $H\geq\Delta$.

The Hamiltonian for the THAFL in our studies reads: 
\begin{eqnarray}
{\cal H}&=&\sum_{i=1} [J_{\|}({\bf S}_{1,i}\cdot {\bf S}_{1,i+1}
+{\bf S}_{2,i}\cdot {\bf S}_{2,i+1})\nonumber\\
&+&J_{\perp} {\bf S}_{1,i}\cdot {\bf S}_{2,i}
-H (S_{1,i}^z+S_{2,i}^z)], ~~~~H>0
\label{hamiltonian}
\end{eqnarray}
where ${\bf S}_{n,i}$ denotes a $S$$=1/2$ spin operator at the 
$i$-th site of the   $n$-th chain.
$J_{\|,\bot}$ are the intra- and inter-chain couplings, respectively. 
To confront the experimental findings
on CuHpCl, we set $J_{\|}$$=1$,
$J_{\perp}/J_{\|}$$=5.28$.

The TMRG technique we adopt  here is implemented in the thermodynamic limit
 and can be used to  evaluate very accurately the  thermodynamic quantities
\cite{wang,xiang} as well as imaginary time auto-correlation 
functions\cite{japs,nwzvl} at very low-$T$ for quasi-1D systems.
Technical aspects of this method can be found in Ref.\cite{xiang}.
In our calculations, the number of kept optimal states $m=200$, while the width
of the imaginary time slice  $\epsilon=0.05$ are used in most   cases.
We have also used different $m$ and $\epsilon$    to verify the accuracy 
of calculations.
The physical quantities presented below are usually 
calculated down to $T\leq 0.02$ (in units of $J_{\|}$). The lowest temperature reached is $T=0.005$. 
The relative errors, being different for different quantities,
are usually much less than, at most about, one percent for derivative quantities at very low temperatures.

We first  determine the spin gap by fitting the spin susceptibility $\chi$,
using the asymptotic formula proposed in \cite{Troyer94}:
$\chi$$=$$A$${\rm e}^{-\Delta/T}/{\sqrt T}, ~~T\rightarrow 0,
$ based on the  quadratic dispersion  with a gap $\Delta$ for the single magnon branch.
Fitting numerical results $\chi$ in the range $T\in [0.168,1]$\cite{Hm}, 
we obtain $\Delta$$=$$4.385$ which is very close to 
the value $\Delta$$=$$4.382$ obtained using the $T$$=$$0$ DMRG method
(with 250 states kept and extrapolated to infinite size).

\begin{figure}
\vspace{-1.0cm}
\epsfxsize=2.5 in\centerline{\epsffile{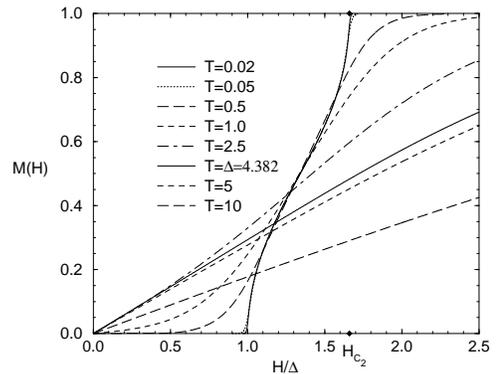}}
\caption{Magnetization  {\it versus} $H$ for different $T$} 
\label{fig3}
\end{figure}

%\noindent ($m$$=$$250$ states were kept and the extrapolation to infinite size was made).

\begin{figure}
\epsfxsize=3.1 in\centerline{\epsffile{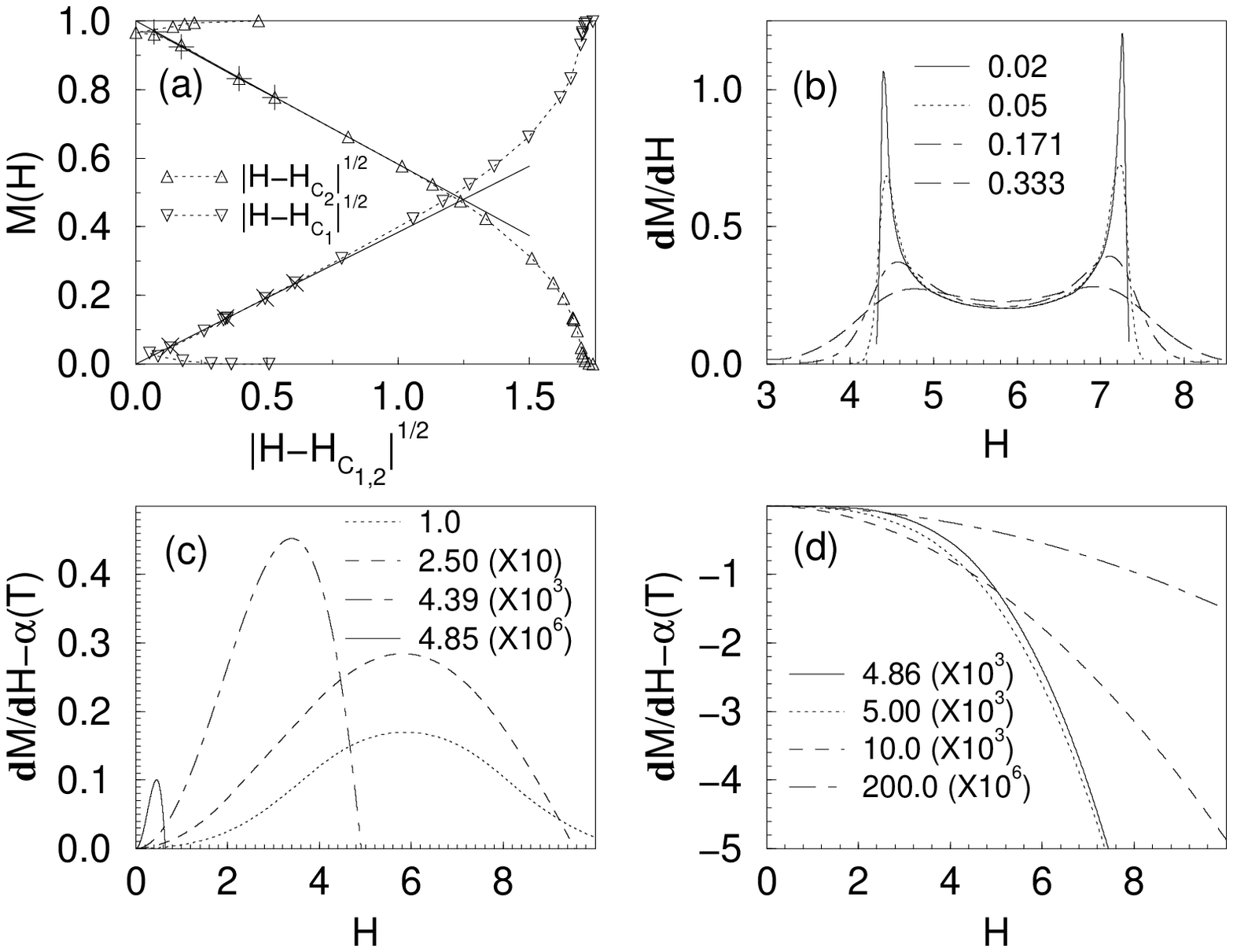}}
\caption[]{(a) Solid lines for the asymptotic behavior Eq. (\ref{mh}). 
Symbols $\times(+)$ for $T$$=$$0$ at $H$$=$ $H_{C_1}^+$ $(H_{C_2}^-)$ 
and up(down) triangles for $T=0.02$ showing the deviation from Eq. (\ref{mh}).
 (b) ${\rm d}M/{\rm d}H$ {\it vs} 
 $H$ (in legends) for $T$$\leq$$T_{0}$; 
 (c) For $T_{0}$$<T$$\leq T_{C}$; 
 (d) For $T$$>$$T_{C}$. 
  In (c) and (d), a constant $\alpha(T)$ is subtracted and derivatives are amplified as 
      seen in parentheses.}
 \label{fig4}
\end{figure}

In Fig. \ref{fig3}, we show the magnetization curves for different values of $T\in[0.02,10]$. 
For quasi-1D spin-gapped systems, it is well-known that at $T=0$, $M=0$ for $H$$\leq$
$H_{C_1}$$=$$\Delta$; $M=1$ 
(in units of $S$) for $H$$\geq$$H_{C_2}$$=$$J_{\bot}+2J_{\|}$\cite{Chaboussant97a} and 
$0<M<1$ for $H\in(H_{C_1},H_{C_2})$.  When $T$$\neq$$0$, $M$ is nonzero for any $H$. 
However, the  critical behavior of the magnetization 
in the vicinity of $H_{C_1}$ and $H_{C_2}$ can only be seen at very low temperatures. 
%and it is  washed out for higher temperatures. 
The  behavior of $M(H)$ is elucidated in Fig. \ref{fig4}. 
First consider   the $T$$=$$0$ case   in (a). 
We have calculated $M(T$$=$$0)$ at $H=7.275,7.25,7.125,700$ and $4.4,4.5,4.625,4.75$
for the upper and lower critical points, respectively. The calculations for $M(T)$
were done with $m$$=$$256$ down to $T$$\approx$$0.005$ (needed) for 
extrapolation to $T$$=$$0$ limit.
Then fitting $M(T$$=$$0)$ at these $H$ gives the following asymptotic form:
\begin{equation}
M(H)=\left\{\begin{array}{ll} 0.380\sqrt{H-H_{C_1}} ~~~&{\rm for}~H=H_{C_1}^+\\
                            1-0.431\sqrt{H_{C_2}-H}~~~&{\rm for}~H=H_{C_2}^-,
     \end{array}
                     \right.\label{mh}
\end{equation}
\noindent  in agreement  with universal square-root singularities of magnetization in gapped 
systems\cite{gn} (see also \cite{chitra,sakai}).
 Independently, we obtain $H_{C_1}=4.3823$ which is even more 
accurate than the value obtained from fitting $\chi(T)$.

When $T$$>$$0$, depending on  the behavior of ${\rm d}M/{\rm d}H$, 
there are three different cases: 
1) ${\rm d}M/{\rm d}H$ has a two-peak structure shown in Fig. \ref{fig4}(b) 
for $T$$<$$T_0$$=$$0.59^{+0.04}$ (positive(negative) numbers in
super(sub)-scripts are bounds of errors\cite{Hm}),
 similar to the $T$$=$$0$ case. 
2) It has a single peak structure at $H$$\neq$$ 0$ in Fig. \ref{fig4}(c) for $T_0\leq T$$<$$T_C
$$=$$4.86_{-0.05}$.
3) It reaches a maximum exactly at $H=0$ for $T$$\geq$$ T_C$ in Fig. \ref{fig4}(d).
Suppose  $\gamma$ is the coefficient of the cubic term 
in the low-$H$ expansion of magnetization, $T_C$ is given  by  $\gamma(T_C)=0$.

\begin{figure}
\epsfxsize=2.5 in\centerline{\epsffile{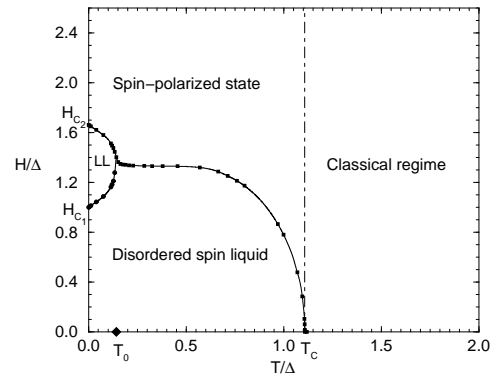}}
\caption{Magnetic phase diagram: the dots fitted by lines as phase boundary in-between the LL, 
the spin-polarized phase and the disordered spin liquid, indicate 
 values of $H$ and $T$  maximizing ${\rm d}M/{\rm d}H$.}
\label{fig2}
\end{figure}

Based on the above observations, we can construct a magnetic phase diagram as shown in Fig. \ref{fig2}. 
Strictly speaking, quantum phase transitions
take place only at $H_{C_{1,2}}$ for $T=0$, and ``phase  boundaries''  are just crossover lines for $T>0$.
At $T=0$: 1) as $H< H_{C_1}=\Delta$, the band edge  of the continuum  has $S^z_{total}=1$, 
with an effective gap $\Delta_{eff}^{<}=H_{C_1}-H$. The ground state
is a disordered spin liquid and thermodynamic quantities decay exponentially at low $T$; 
2) As $H_{C_1}\leq H\leq H_{C_2}$, the gap vanishes and we find  a
range of linear in $T$ dependence for the specific heat and   finite
values for the susceptibility, which is characteristic for the Luttinger liquid (LL);
3) The ground state becomes fully polarized when $H>H_{C_2}$.
 There  thermodynamic quantities again decay
exponentially with $\Delta^{>}_{eff}=H-H_{C_2}$ at low $T$.
When $T>0$, the LL regime shrinks gradually and disappears at $T=T_0$ and $H=H_m$, 
beyond which the system ``forgets'' about $H_{C_1}, 
H_{C_2}$ and $\Delta$\cite{ners}. 
The other  two phases continue to exist  until $T=T_C$, and 
one finds  a classical regime for $T>T_C$.

Compared with the phase diagram proposed on the basis of  $1/T_1$ measurements\cite{Chaboussant98}, 
the major difference is the absence of the ``quantum critical'' phase in Fig. \ref{fig2}.
We note that the requirement for exhibiting the universal ``quantum critical'' 
behavior $ J \gg \Delta$ \cite{hertz} is not satisfied in our case. 
In addition, for the classical regime, we found $T_C=1.109_{-0.011}\Delta$ instead of $T_C=\Delta$.
In Ref. \cite{Chaboussant98}, the phase boundaries for the quantum critical phase with two gapped phases
are given as the onset of the exponential behavior in  $1/T_1$ at $T=\Delta^{>,<}_{eff}$.
(Consequently, one  obtains $H=H_{C_{1,2}}$ at $T=0$ and $T_C=\Delta$ from 
$\Delta^{<}_{eff}=0$ at $H=0$.) 
The divergence of $1/T_1$ presumably disappears at the boundary of the LL regime. 
In fact, $1/T_1$ has contributions from  magnon scattering with momentum transfer  of both $q=0$  
and $q=\pi$. The former process corresponds to a larger  gap in the continuum, 
but contributes substantially to the relaxation\cite{felix}.
In our calculations the critical behavior of $M(H)$ defining the quantum phase transitions
 is identified directly at $T=0$. 
When $T>0$, a straightforward  extension of this definition
gives rise to all regimes except for the ``quantum critical'' phase. 
%As discussed below, the interesting features of several physical quantities
%including $1/T_1$ can be interpreted within our phase diagram Fig. \ref{fig2}.
We should also  mention that when $J_{\bot}\rightarrow0$, one has $\Delta\rightarrow0$, $T_0\rightarrow T_C$
and $H_m\rightarrow 0$.

\begin{figure}
\epsfxsize=2.5 in\centerline{\epsffile{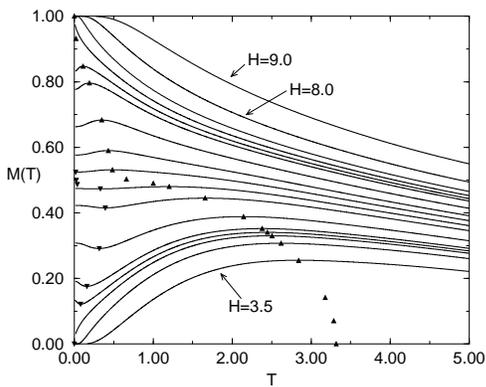}}
\caption{Magnetization {\it vs.} temperature: curves for $H$ $=$ $9,$ $8,$ $7.415,$ $7.28,$ $7.125,$ 
$7,$ $6.75,$ $6.25,$ $6,$ $ 5.75,$ $5.5,$ $5,$ $4.625,$ $4.5,$ $4.385,$ $4.125,$ $3.5$
from top to bottom. Symbols $\bigtriangleup$ ($\bigtriangledown$) denote   maxima (minima) at each $H$. 
Extra points added between curves are 
for maxima at $7.25,$ $5.875,$ $5.812,$ $2,$ $1,$ $0.005$ and minima
 at $5.875,$ $5.812$.}
\label{fig5}
\end{figure}

Now elaborate more on the temperature dependence
of $M(T)$ for various given $H$ as shown in Fig. \ref{fig5}. 
Consider a cooling process. For $T>T_C\approx\Delta$, $M$ monotonically but slowly increases 
for all $H$, whereas it can change non-monotonically depending upon the values of $H$
for $T<T_C$. When $H$$>$$H_{C_2}$, $M$ continues to increase and saturates exponentially
 with $\Delta_{eff}^{>}$.
However, when $H$$<$$H_{C_1}$, $M$   first goes  up, then down and finally 
decays to zero exponentially with $\Delta_{eff}^{<}$.
When $H_{C_1}<H<H_{C_2}$, there is always a maximum at $T\neq 0$,
 close to the boundary between the polarized phase and 
 LL for $H\geq H_m\approx 6$\cite{Hm},
while separating the former from  the disordered spin liquid phase elsewhere. 
There is also a minimum  for   LL. The positions of minima are 
at $T=0$ for $H\geq H_m$, otherwise  they are close to the boundary of the LL regime.

The origin of this nontrivial temperature dependence, in particular,
the presence of minima and maxima at low temperatures, is not fully understood.
One possible interpretation is due to the excitation spectrum in the presence of MF.
$S^z_{total}$ is  a good quantum number for the Hamiltonian 
 (\ref{hamiltonian} ) and thus
different energy bands are shifted by $-S^z_{total}H$.
When $H< H_{C_2}$, the ground state is not a fully polarized state(FPS),
but the low-lying excitations correspond to positive $S_{total}^z$, and
the maximum, roughly speaking, corresponds to the ``maximally
polarized" state. This is true so far $H$$>$$H_m$. 
%Moreover, the saturation for $H$$>$$H_{C_2}$ and the decay for $H$$<$$H_{C_1}$,
% respectively, reflect two kinds of low-lying spectra:
%the fully polarized ground state and dominantly polarized low-lying excitations;
%the non-magnetic ground state and weakly polarized low-lying excitations.
% When $H$ decreases from $H_{C_2}$, a peak appears instead of   saturation,
%because the ground state is still a dominantly polarized state but no longer a 
%fully polarized state (FPS). Then the FPS should be
%related to the position  of the main peak at each $H$ (the low-lying excitations
%within the range $T_{peak}$ are ``more'' polarized  states than the ground
%state for $H$$>$$H_m$). The sharpness of the peak depends  essentially on 
%the density of states and the corresponding $S_{total}^z$ near  FPS
%in the spectrum. Those states near    FPS are dominantly polarized 
%with $S_{total}^z>0$ for $H>H_m$\cite{Hm}, but can be weakly 
%  polarized states or have $S_{total}^z\leq0$ when $H<H_m$. 
On the other hand,  the minimum  at $T$$\neq$$ 0$  originates from low-lying 
states in $S_{tot}^z$$=$$0$ subspace. 
Those states intersect  with  FPS at $ H = H_m$.
Notably, $H_m$ also corresponds to an extrapolation of the boundary between the
two gapped phases in the phase diagram ( Fig.\ref{fig2}).  It is also curious
to note that in Fig. \ref{fig5}, the curves are roughly symmetric w.r.t. $M=0.5$,
if we focus on the low temperature part.  This reflects the particle-hole
symmetry of the problem in the fermion representation\cite{chitra,Chaboussant98}.

\begin{figure}
\epsfxsize=3.0 in\centerline{\epsffile{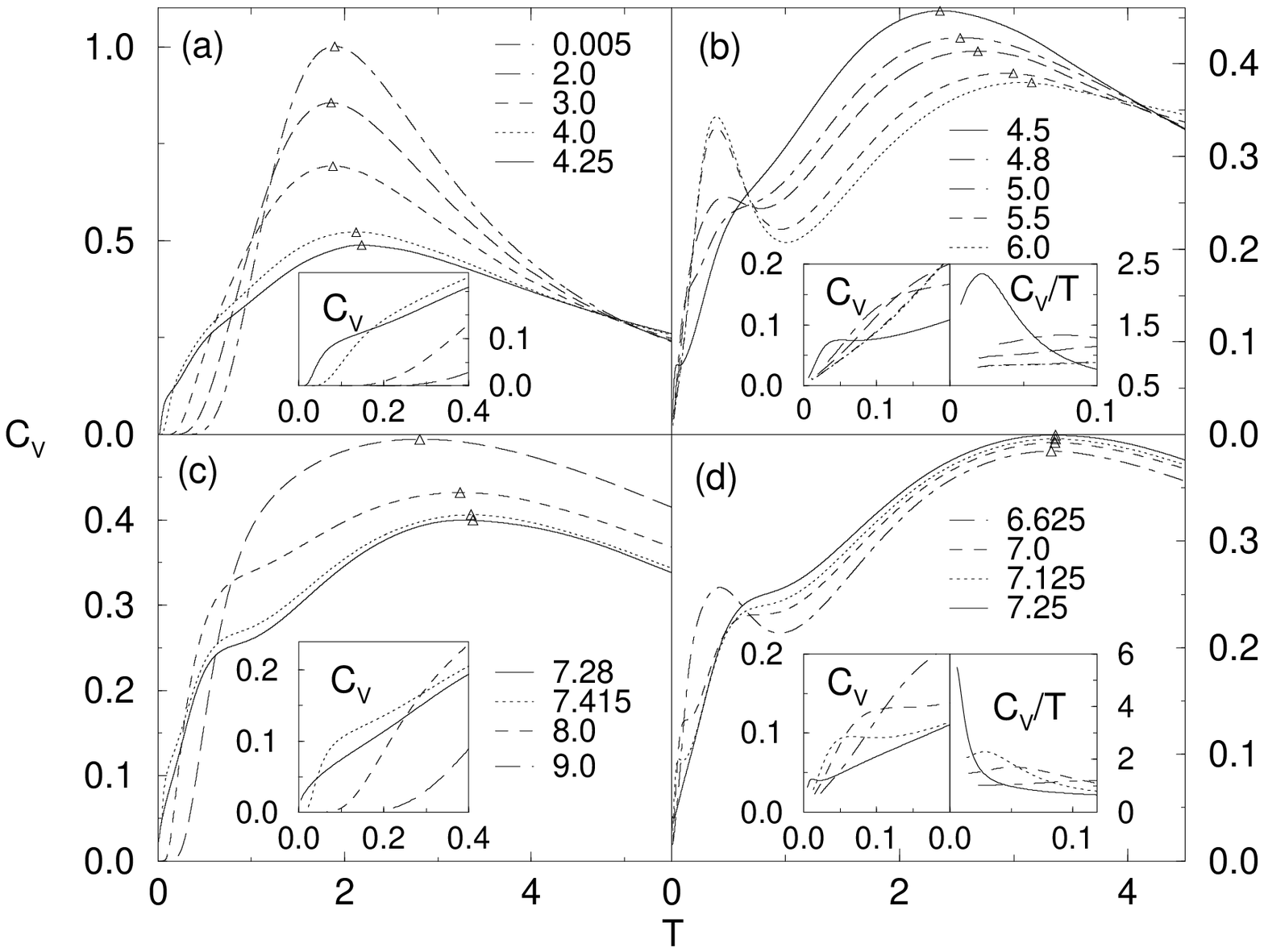}}
\caption{Specific heat at various $H$ (legends): For $H<H_{C_1}$ in (a);
 $H_{C_1}<H<H_{C_2}$ in (b) and (d); $H\geq H_{C_2}$ in (c). Insets for low-T behavior: $C_v$ in 
(a)-(d) and $C_v/T$ in (b) and (d). Triangles denote 
maximum $C_v$.}
\label{fig6}
\end{figure}

We now turn to the specific heat which, 
similar to $M(T)$, shows different  behavior depending on $H$.
When $H<H_{C_1}$ in Fig. \ref{fig6}(a), $C_v$ has a single peak structure as 
expected. The MF reduces $\Delta_{eff}^{<}$, 
and changes dramatically the line shape near
$T=0.5$,  as $H\rightarrow H_{C_1}^-$. This is a signature of approaching the
quantum critical point\cite{hertz}.
When $H_{C_1}$$<H$$<H_{C_2}$ in Fig. \ref{fig6}(b) and (d), a 
second peak at low $T$ is developed 
exhibiting the LL  behavior. Linear-$T$ dependence
 is shown in the insets of (b) and (d). 
Moreover, at $H=H_{C_1}^+$, the cusp still remains and 
at $H=H_{C_2}^-$ a shoulder emerges. When $H\geq H_{C_2}$ in Fig. \ref{fig6}(c), 
the shoulder can still be seen for $H=H_{C_2}^+$, although the second peak vanishes.
At low-$T$, $C_v$ decreases exponentially with $\Delta_{eff}^{>}$.
We note the cusps 
and shoulders appear  outside the LL regime but at the vicinity of its boundary.
For those $H$ at which a larger second peak shows up, the local minima are
also located  outside the LL regime,  but nearby.

\begin{figure}
\epsfxsize=2.3 in \centerline{\epsffile{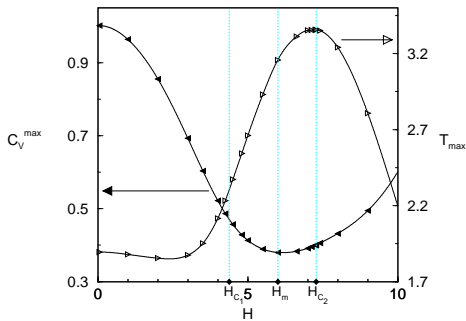}}
\caption{The maximum specific heat $C_v^{max}$ (to the left) and the corresponding 
temperature $T_{max}$ (to the right) {\it versus} $H$.}
\label{fig7}
\end{figure}

It is also instructive to see the MF effects on the maximum specific 
heat $C^{max}_v$ and 
corresponding temperature $T_{max}$. As seen in Fig. \ref{fig7}, when $H$ is 
applied, $C_v^{max}$ first declines
as a gradual response to the splitting. At $H=H_m$, because of the crossing of two kinds 
of energy states discussed above, $C^{max}_v$ arrives at a minimum. 
Moreover, we surprisingly found that the curvature of $T_{max}$ 
changes its sign at $H_{C_1}$, while it reaches a maximum at $H=H_{C_2}$. 
These interesting features  show an intrinsic aspect of the MF effects
as demonstrated  consistently by the magnetization and the specific heat.

Finally we discuss the bearings of our numerical results
on experimental findings  of  a diverging NMR relaxation rate 
and peculiar specific heat  behavior.
Recently, 
Chaboussant {\it et al.} found that the NMR rate anomalously increases 
for $H_{C_1}$$\leq$$ H$$\leq $$H_{C_2}$ when $T$ decreases\cite{Chaboussant98}. 
These authors  attributed this anomalous increase  to quasi-1D behavior.
Our results  show that it is  indeed  an intrinsic MF effect on the spin ladders.
In fact, the increasing of the NMR rate starts already at the vicinity of the LL  regime
which is bounded by $T$$\approx$$ 1.6K$. This should share the same physical origin 
as the occurrence of the cusps and shoulders for $C_v$ at the vicinity of the LL regime. 
On the other hand, Hammar {\it et al.}\cite{Hammar98} have measured 
$C_v$ up to $H=9$T, which is about $(5.10\pm 0.17)J_{\|}$, taking into account
the difference  between the experimental and numerical results 
for the gap due to other interactions\cite{Hawyard96,Chaboussant98}. 
As seen in Fig. 3 of Ref. \cite{Hammar98}, the overall feature is consistent with 
our results in Fig. \ref{fig6}(a) and (b), {\it e.g.}   
the shift of $C_v^{max}$ and $T_{max}$ as well as 
the abrupt change  for $H$$=$$6.6$T. The development of the shoulder and the second peak
at low temperatures is clearly seen in more recent measurements
\cite{Cal} in full agreement with our calculations. When $H_{C_1}$$<$$H$$<$$H_{C_2}$, 
we notice that a narrow subpeak emerges from the second peak
at lower temperature (see Fig. 2(b) of Ref. \cite{Cal}). 
The subpeak might indicate the on-set of the 3D effects \cite{Hammar98,Cal}, 
while the second peak represents the magnetic effects on the THAFL.
On contrary, the NMR rate changes smoothly {\it versus} temperature in the LL regime. Therefore,
the anomalous behavior of the NMR rate results from the magnetic effects of THAFL
as a characteristic feature of quasi-1D gap systems.

In conclusion, we have proposed a magnetic phase diagram for the two-leg ladders.
We emphasize that most of the striking MF effects show up in the LL regime, 
giving rise to the divergence of the NMR rate and anomalous specific heat observed  in  experiments. 
Moreover, this magnetic phase diagram is generically valid  for other  spin-gapped 
systems as well  and the results on $M(T)$ should also shed some  light on other
 quasi-one dimensional fermion  systems which involve either a charge gap or a band gap,
with $H$ replaced by the chemical potential.
% $\mu$. 

We are grateful to D. Loss, A.A. Nersesyan, T.K. Ng,  B. Normand and Z.B. Su
for fruitful discussions and to  G. Chaboussant for helpful correspondence.


\begin{references}
\bibitem{Hm} The detail will be discussed elsewhere.
\bibitem{ners} In Matsubara Green functions gap $\Delta$ is compared with $2\pi T$,
and $\Delta / 2\pi \approx 0.7$,  very close to $T_0$ in our case. This 
remark is due to A.A. Nersesyan.
\end{references}
\end{document}